# Comment on "Memory Effects in an Interacting Magnetic Nanoparticle System"

R. K. Zheng and X. X. Zhang[*]

*Department* evidenced *of Physics and Institute of Nano Science and Technology, The Hong Kong University of Science and Technology, Clear Water Bay, Kowloon, Hong Kong*

Recently, Sun *et al* reported that striking memory effects had been clearly observed in their new experiments on an interacting nanoparticle system [1]. They claimed that the phenomena evidenced the existence of a spin-glass-like phase and supported the hierarchical model. No doubt that a particle system may display spin-glass-like behaviors [2]. However, in our opinion, the experiments in Ref. [1] cannot evidence the existence of spin-glass-like phase at all. We will demonstrate below that all the phenomena in Ref. [1] can be observed in a *non-interacting* particle system with a size distribution. Numerical simulations of our experiments also display the same features.

Co particles of 5 nm in diameter synthesized by chemical method were used in our experiments. To avoid the interparticle interaction, Co particles were dispersed into hexane very dilutedly. We followed the measurement procedures in Ref. [1] to inspect the memory effects. ZFC and FC magnetizations were measured in a 100 Oe field with a temperature sweeping rate of 2 K/m, shown in Fig. 1 (a). For the T$\downarrow$ measurement, the cooling was suspended for 2 h at 25 and 15 K, respectively, where the field was cut off. The T$\downarrow$ curve displays step-like feature. When the system is heated back, The T↑ curve almost repeats the T↓ one, i.e., exhibits memory effects. We also reproduced the features in Fig. 3 and Fig. 4 of Ref. [1] in the non-interacting Co particle system. However, a non-interacting particle system should not display spin-glass-like behaviors. To manifest that the non-interacting Co particle system is not spin-glass-like, we performed low frequency ac susceptibility and relaxation measurements following the methods in Ref. [3]. Nevertheless, memory effects of spin-glass were not observed (not shown here). Therefore, the non-interacting nanoparticle system is not spin-glass-like. We will show that the memory effects observed by the methods of Sun *et al* are merely a superparamagnetic relaxation effect, nothing to do with spin-glass. Since each single domain particle has its own blocking temperature $T_B$, a particle system has a $T_B$ distribution (or spectrum). Particles with different $T_B$ contribute to the relaxation at different temperature range. We extracted parameters from experimental data to simulate the memory effects using magnetic relaxation theory. The simulated curves shown in Fig. 1 (b) also display the same memory effects as experiments. Therefore, the memory effects are simply a magnetic relaxation effect. Before the stop at 25 K, the T↓ curve coincides with FC. After the field is cut off and a long time wait (2 h), magnetization significantly relaxes from "A" to "B". When the field is switched on again, magnetization can only return to "C" within a short time of one measurement (tens of seconds). The nanoparticles away from their equilibrium states (the states in FC) are responsible for the difference between "A" and "C". The difference is kept at lower temperature, because relaxation dramatically slows down as temperature decreases. The nanoparticles in their equilibrium states behave the same as in FC, so the T↓ curve presents similar trend with FC. The stop at 15 K can also be understood in the same way. When the system is heated back, the T↑ curve almost merges the T↓ one before stop temperature (15 or 25 K) because the nanoparticles out of equilibrium still keep their magnetization nearly unchanged. From stop temperature, these nanoparticles significantly relax towards their equilibrium states (FC curve). The T↑ curve joins FC very quickly due to the dramatically speedup of relaxation at higher temperature.

---

[*] E-mail:phxxz@ust.hk

The only mechanism responsible for the memory effects is superparamagnetic relaxation. Therefore, the experiments presented in Ref. [1] cannot evidence the existence of spin-glass. The relaxation of individual nanoparticles is exponentially fast, however, a nanoparticle system with an energy barrier distribution usually relax quite slowly (e.g., logarithmic), which can mimic the slow dynamics of spin-glass. Therefore, Special caution should be paid in studying nanoparticle systems.

[1] Y. Sun, M. B. Salamon, K. Garnier, and R. S. Averback, Phys. Rev. Lett. **91**, 167206 (2003).

[2] T. Jonsson, P. Svedlindh, and M. F. Hansen, Phys. Rev. Lett. **81**, 3976 (1998); and reference therein.

[3] K. Jonason, E. Vincent, J. Hammann, J. P. Bouchaud, and P. Nordblad, Phys. Rev. Lett. **81**, 3243 (1998).

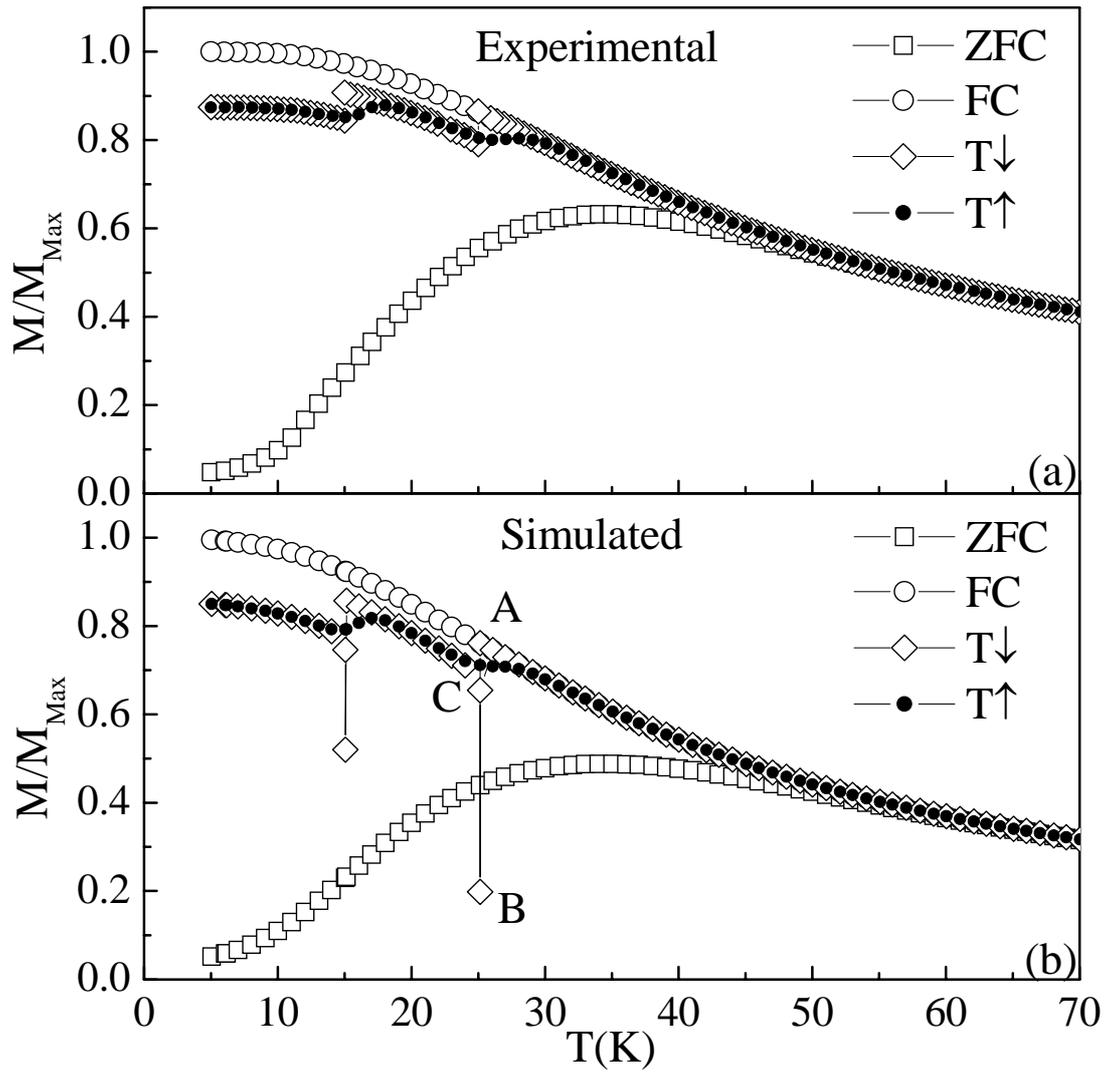

Fig. 1 (a) Non-interacting Co nanoparticles show "memory effects" of Ref. [1]. (b) Simulations of experiments.